\documentstyle[prl,aps,epsf,floats,twocolumn]{revtex}

\def\boldhat#1{{\bf{\hat #1}}}

\begin{document}

\advance\textheight by 0.5in
\advance\topmargin by -0.25in
\draft

\twocolumn[\hsize\textwidth\columnwidth\hsize\csname@twocolumnfalse%
\endcsname

\title{Communication through a Diffusive Medium: \\
Coherence and Capacity}
\author{Aris L. Moustakas${}^1$, Harold U. Baranger${}^{1,2}$, \\
Leon Balents${}^{1,3}$, Anirvan M. Sengupta${}^1$
and Steven H. Simon${}^1$}
\address{ ${}^1$Bell Labs, Lucent Technologies, 700 Mountain Avenue, 
Murray Hill, NJ 07974}
\address{ ${}^2$Department of Physics, Duke University, Durham, NC 27708--0305}
\address{ ${}^3$Department of Physics, University of California, Santa
  Barbara, CA 93106}



\maketitle


\begin{abstract}
  Coherent wave propagation in disordered media gives rise to many
  fascinating phenomena as diverse as universal conductance
  fluctuations in mesoscopic metals and speckle patterns in light
  scattering. Here, the theory of
  electromagnetic wave propagation in diffusive media is combined with
  information theory to show how interference
  affects the information transmission rate between antenna arrays. 
  Nontrivial dependencies of the information capacity on the
  nature of the antenna arrays are found, such as the dimensionality
  of the arrays and their direction with 
  respect to the local scattering medium.
  This approach provides a physical
  picture for understanding the importance of scattering in 
  the transfer of information through wireless communications.
\end{abstract}

\vspace*{5pt}
] 

\newpage

The ongoing communications revolution has motivated researchers to
look for novel ways to transmit information {\em (1, 2)}.  One recent
development {\em (3, 4)} is the suggestion that, contrary to long-held
beliefs, random scattering of microwave or radio signals may enhance
the amount of information that can be transmitted on a particular
channel.  Prompted by this suggestion, we introduce a realistic
physical model for a scattering environment and analytically evaluate
the amount of information that can be transmitted between two antenna
arrays for a number of example cases.  On the one hand, this lays a
new foundation for complex microwave signal modeling, an important
task in a world with ever-increasing demand for wireless
communication, and, on the other, introduces a new arena for
physicists to test ideas concerning disordered media.

{}From information theory {\em (5)}, the capacity of a channel between
a transmitter and a receiver, that is, the maximum rate of information
transfer at a given frequency, can be described in terms of the
average power of the signal $S$ and the noise $N$ at the receiver: $C
=\log _{2}(1 \!  + \! S/N)$. More generally {\em (2)}, the
communication channel connecting several transmitters and receivers is
described by a matrix $G_{i\alpha }$ giving the amplitude of the
received signal $\alpha$ due to transmitter $i$. The information
carried by the channel can be characterized using several quantities,
such as the capacity or mutual information, which are typically
functionals of the matrix $G$, which must be known in order to predict
these quantities.  Often $G$ cannot be predicted for actual systems,
such as wireless communication networks or optical fibers, because of
the complicated scattering and interference of waves that is involved.
It is crucial, therefore, to develop physical models for the signal
propagation, as it is only through such models that one can understand
the real effects of scattering and interference on the amount of
information that can be communicated.

In many cases, only partial information is available for prediction;
in these situations, one only has a statistical description of $G$.
Instead of making assumptions about $G$ directly, the usual procedure
in information theory, we introduce statistical models for the physical
 environment from which we  derive the properties of
$G$. The advantage of this procedure is that simple physical models
can yield very non-trivial properties of $G$.

Statistical descriptions of the environment have been quite successful
in the physics of disordered media {\em (6 -- 8)}. The simplest of
these is diffusive propagation.  In our case of electromagnetic
propagation in the context of wireless communication, diffusion is
known to work well in various circumstances {\em (9)}, and
simple extensions seem relevant for many others. From a diffusive
approach, one finds the moments of the distribution of $G$. These will
enable us to calculate information-theoretic quantities (for example the
capacity) using a replica field theory approach to random matrix
theory {\em (10)}.  Implicit in this approach is the 
assumption that the full distribution of $G$ is sampled, which is
realistic in many real world situations where the environment is
changing.  However, when the number of antennas is large, many quantities of
interest become strongly peaked around their average and this
assumption can be relaxed. 

In a statistical description, the scattering of the signal is
characterized by the mean-free path, $\ell$, corresponding roughly to
the distance between scattering events. When $\ell$ is large compared
to the wavelength $\lambda $ but small compared to the distance $d$
between the two arrays, the wave propagation becomes diffusive {\em
(7, 8)}.  This has been analyzed in the past in the context of
electron diffusion in metals {\em (6)} and light propagation in solids
{\em (7, 11)}. In the case of wireless propagation, with signals in
the 2 GHz region, $\lambda \sim $ 10-15 cm, while $\ell $ is of order
meters for indoors and tens of meters for outdoors propagation, so
diffusion is applicable.

In the diffusive regime $\lambda \ll \ell $, to leading order in
$\lambda /\ell $, only the quadratic correlations $\langle G_{i\alpha
}^{{}}G_{j\beta }^{*} \rangle$ are non-negligible and therefore
describe the system, where the brackets represent an average over
realizations of the disorder.  Higher cumulants of $G$ are of higher
order in $\lambda /\ell $. Therefore, the distribution of $G$ is
Gaussian with zero average {\em (6 -- 8)}. The leading term in
$\langle G_{i\alpha }^{{}}G_{j\beta }^{*}\rangle$ is evaluated by a
summation of so-called ladder diagrams {\em (7)} corresponding to
processes in $\langle G_{i\alpha }^{{}}G_{j\beta }^{*}\rangle$ where
the waves from antennas $i$ to $\alpha $ and from $j$ to $\beta $
propagate through the scattering medium along identical paths except
for segments of order $\ell$ at each end.

In several realistic situations discussed below, the correlations take
a particularly simple form:
\begin{equation}
\langle G_{i\alpha } G_{j\beta }^{*}\rangle=R_{\alpha \beta }\
\frac{S}{n_{T}}\ T_{ij} \;. 
\label{RST}
\end{equation}
Here $R_{\alpha \beta }$ and $T_{ij}$ are matrices describing the
correlations of the receiving and transmitting antennas, respectively,
and $S = {\rm Tr}\{GG^{\dagger }\}/n_{R}$ is the average power
received at each of $n_R$ receivers assuming independent signals from
each of $n_T$ transmitters. $R$ and $T$ are normalized such that ${\rm
  Tr}\{T\} \!= \!n_{T}$ and ${\rm Tr}\{R\} \!= \!n_{R}$.  The
factorization in Eq.  \ref{RST} of the receiver and transmitter
information reflects the dominance of the ladder terms: only the
segments near the antennas contribute a net phase, so only local
information about the antennas is relevant.  Furthermore, under
general circumstances, $T$ can be expressed in terms of the response
$\chi_i (\boldhat{k},\boldhat{\epsilon})$ of antenna $i$ to an
incoming plane wave in direction $\boldhat{k}$ with polarization
$\boldhat{\epsilon}$ and a weight function
$w(\boldhat{k},\boldhat{\epsilon})$ giving the fraction of incident
power in that direction with that polarization:
\begin{equation}
T_{ij} \propto \int \!\! d\boldhat{k} \sum_{\boldhat{\epsilon}}  \,\,  
w(\boldhat{k},\boldhat{\epsilon}) 
\chi_i (\boldhat{k},\boldhat{\epsilon}) 
\chi_j^* (\boldhat{k},\boldhat{\epsilon}) \;.
\label{chichi}
\end{equation}
$R$ is defined similarly.

Having summarized the coherent diffusion results that we will need, we
now turn to considering information-theoretic quantities.  While there
are many such quantities that one could study with the diffusive
techniques outlined above, for concreteness we focus on a problem of
relevance to wireless communication
applications {\em (2 -- 4, 12, 13)}.
We assume that: (a) the receiver knows $G$, but the transmitter does
not, (b) the noise at each receiving antenna has the same power $N$
and is Gaussian uncorrelated, and (c) after a long time, the entire
space of possible $G$ matrices is explored as the environment changes,
subject to the statistical properties of Eq.  \ref{RST}.  Under these
circumstances, the maximum time averaged information transfer rate is
given by the so-called capacity, formally defined in {\em (14)}, given by
{\em (2 -- 4)}
\begin{equation}
C = \max_{Q} \left\langle {\rm Tr}\left\{ 
\log _{2}[I_{n_{R}}+\case{1}{N} \,\, G\ Q\ G^{\dagger }]   \right\} \right\rangle
\label{matrixshannon}
\end{equation}
where $I_{n_{R}}$ is the $n_{R}\times n_{ R}$ unit matrix and $Q$ is
the non-negative-definite $n_{T} \times n_{T}$ covariance matrix
describing the correlations between the signals from the transmitter
antennas.  Here, the angled brackets again indicate an average over
realizations of $G$ with correlations defined by Eq. \ref{RST}.  With
the normalization of $G$ given above, the constraint on the maximum
power transmitted yields the condition ${\rm Tr}[Q] \le n_{T}$.  For a
fixed covariance $Q$, the maximum information transfer rate is given
by the same expression only without the $\max_Q$ in front.
Maximization over $Q$ then gives the capacity of the channel.

The assumptions (a)-(c) given above are appropriate for certain real
world systems, and have been previously studied {\em (3, 4)}.
Assumption (a) is appropriate if the transmitter occasionaly sends
known signals so the receiver can determine $G$, and if $G$ changes
slowly enough that the channel needs to be probed only very
occasionally.  We also assume that the receiver cannot send
information about $G$ back to the transmitter.  Assumption (c), while
necessary to make Eq.  \ref{matrixshannon} formally correct, is not
too important in practice for systems with a large number of antennas
where the maximum information transfer rate for a given $Q$ is
strongly peaked about its average value {\em (3, 15)}.

An important implication of Eq. \ref{matrixshannon}, recently pointed
out in {\em (3, 4)}, is that if the environment is sufficiently
multipath, the capacity increases linearly with the number of antennas
even when the total transmitted power is fixed, greatly exceeding the
capacity of line-of-sight propagation.  This result may seem
surprising since scattering may reduce the received signal.  However,
the scattering also produces many independent paths which interfere at
both the transmitter and receiver. This interference can be exploited
by placing several antennas at both ends in order to increase the
capacity.  We now illustrate this dramatic result with a simple
example.

\begin{figure}[tbp]
\epsfxsize=3.3in\epsffile{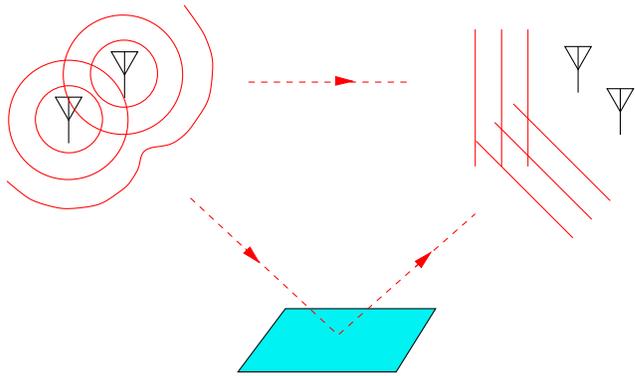}\vspace{0.1in}
\caption{ Transmitter and receiver antenna arrays in the presence of a
single scatterer (mirror). Electromagnetic waves interfere at both
transmitter and receiver antennas. The presence of the scatterer introduces
an additional path of wave propagation between the two arrays and thereby
additional interference at both ends. As a result, the receiver can resolve
the individual signals from the two different transmitting antennas.}
\label{fig1}
\end{figure}

Consider the case of two antennas at each end and assume that a single
scatterer (a mirror) is at some point between the two arrays (Fig.
\ref{fig1}). Here and throughout this paper the Fresnel limit will be
assumed: the size of the transmitter and receiver arrays, is much
smaller than $\sqrt{\lambda d}$.  In the absence of scattering there
is only line-of-sight propagation.  Due to the small effective
aperture of the receiving array, it is impossible to resolve the
different transmitting antennas.  Thus, the capacity is the same as if
there were only one transmitter and one receiver antenna, except that
the noise level is reduced by a factor of 2 since it is averaged over
2 receivers.  Mathematically, this corresponds to $G$ being a rank one
matrix, indicating only one channel of communication.  Including the
mirror, the signal received is now the sum of the direct line-of-sight
and the scattered amplitudes.  If the two incoming waves are at very
different angles, then the aperture of the receiving array is
sufficient to resolve the two waves separately.  It is thus possible
to distinguish the two signals originating from the two
transmitters. Now, $G$ is of rank two so there are two nonzero terms
(two channels) in Eq. \ref{matrixshannon}, roughly doubling the
capacity.

Foschini and Gans argue {\em (3)} that for many scatterers, it is
possible for $G$ to become of full rank {\em (13)} such that the
capacity is proportional to the number of antennas.  An increase in
capacity can thus be obtained even if $S/N$ is reduced by scattering
since this latter effect reduces the capacity only logarithmically.
In order to derive their results, they assume that $G$ is a random
matrix with completely uncorrelated elements.  Their situation (fully
uncorrelated matrix elements) corresponds to $R$ and $T$ being
identity matrices in Eq. \ref{RST} so that the optimal $Q$ in
Eq. \ref{matrixshannon} is identity also.  This yields, in fact, an
upper bound on the capacity which can be realized only in very special
circumstances.

We now generalize the results of Foschini and Gans {\em (3)} to
situations with nontrivial correlations ($R$ and $T$ not identity)
which actually occur in real systems.  The capacity {\em (14)}
defined by Eq.  \ref{matrixshannon} with such nontrivial correlations
has never been considered previously.  These correlations (nontrivial
$T$ and $R$) indicate some redundancy in both the transmitter and
receiver arrays thus reducing the capacity.

Surprisingly, we find that the capacity (Eq. \ref{matrixshannon}) can
be evaluated analytically for a large number of antennas using
replica field theory extensively applied in statistical
physics {\em (10)} once $T$ and $R$ are known (that is given
$\langle G_{i\alpha }^{{}}G_{j\beta }^{*}\rangle$). The result of this
approach is a set of algebraic equations for the capacity in terms of
the eigenvalues of $R$ and $T$ which will be described in detail
elsewhere.  In contrast, attempting to evaluate Eq.
\ref{matrixshannon} numerically by generating realizations of $G$ with
appropriate covariance would be quite difficult since one must then
maximize over $Q$.

We now analyze illustrative situations involving scattering media. The
simplest case is when scalar waves are considered and the antennas are deep 
inside a uniform and isotropic scattering medium  with mean free path 
large compared to the size of the arrays.
 In addition, if we assume that the antennas are ideal,
making a perfect measurement of the field at one point without
distorting it in any way, then $\chi_i (\boldhat{k})= \exp({i k_0
  \boldhat{k} \cdot {\bf r}_i})$ (no $\boldhat{\epsilon}$ appears for
scalar waves).  Due to isotropy, $w(\boldhat{k}) = 1$, and thus using
Eq. \ref{chichi}, we obtain $T_{ij}=f(k_0|{\bf r}_{i}-{\bf r}_{j}|)$
and $R_{\alpha \beta }=f(k_0|{\bf r}_{\alpha }-{\bf r}_{\beta }|)$
with $ f(x)= \sin(x)/x$ where $k_{0}=2\pi /\lambda $ and ${\bf r}_{i}$
are the antenna positions.  When the antennas are separated by
multiples of $\lambda /2$, all antennas of each array on a straight
line, there are no correlations, and $R$ and $T$ are unit matrices as
assumed by Foschini and Gans. Any other configuration has nontrivial
correlations and therefore lower capacity.

\begin{figure}[tbp]
\epsfxsize=3.3in\epsffile{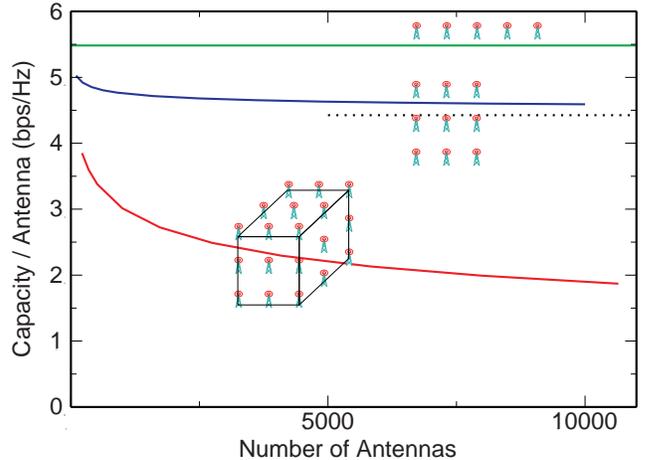}\vspace{0.1in}
\caption{Information capacity per antenna (bits per second per Hertz)
between identical antenna arrays in a uniform isotropic medium as a
function of the number of antennas, for three representative
situations. In all cases the nearest neighbor distance is $\lambda/2$,
the signal to noise ratio is $ 100$, and we have considered scalar
waves. {\bf Top}: Linear arrays of antennas. In this configuration the
antennas are completely uncorrelated, yielding the maximal capacity
per antenna, $C/n=$ 5.483 bps/Hz/antenna for this signal to noise
ratio. {\bf Middle}: Square arrays. The asymptotic value of the
capacity per antenna is predicted to be 4.43 bps/Hz/antenna.  Note the
slow convergence of the capacity per antenna to its asymptotic value;
this effect is due to the slow decay of correlations ($f(x)$) of
antennas at large distances. {\bf Bottom}: Cubic arrays. Here the
capacity per antenna decays as $n^{-1/3}$ up to log corrections since
only the surface layer of antennas contributes to the information
capacity.}
\label{fig2}
\end{figure}

An important consequence of the form of the correlation matrices is
the scaling of the capacity with the number of antennas. In the limit
of a large periodic array of antennas, the eigenvectors of $T$ and $R$
are plane-waves (by Bloch's theorem) so one can find the
eigenvalues analytically. When antennas are placed on a one or two
dimensional array (Fig. \ref{fig2}), the capacity is
proportional to the antenna number. However, when they are placed in a
three dimensional array, the capacity scales as $n^{2/3}$ up to a
logarithm,  not linearly.  The model of a
scattering medium discussed above gives an intuitive interpretation of
this result: a finite thickness of antennas on the surface is
sufficient to resolve the incoming waves from the transmitting
antennas; additional antennas in the interior are redundant.
Therefore the optimal $Q$ is such that less power is transmitted from
these interior antennas.

As a second example we consider ideal dipole antennas and vector
electromagnetic waves.  The average power propagating is the same as
for scalar waves, since vectorial diffusion (non-zero spin) decays
exponentially {\em (11)}. The correlation matrices are now given by
using $\chi_i (\boldhat{k},\boldhat{\epsilon}) = (\boldhat{p}_i \cdot
\boldhat{\epsilon} ) \exp({i k_0 \boldhat{k} \cdot {\bf r}_i})$ in Eq.
\ref{chichi}, where $\boldhat{p}_i$ is the polarization of antenna
$i$.  From the resulting expression, one finds that if antennas are
positioned on a straight line with polarization at an angle $\theta
=54.73^{o}$ ($\cos^2 \theta =\case{1}{3}$) with respect to that line, the
correlations again vanish when antenna separations are multiples of
$\lambda /2$. In this case, then, the ideal situation of Foschini and
Gans is again realized; however, for any other angle, correlations play
a role even if the antennas are separated by half wavelengths.

\begin{figure}[tbp]
\epsfxsize=3.3in\epsffile{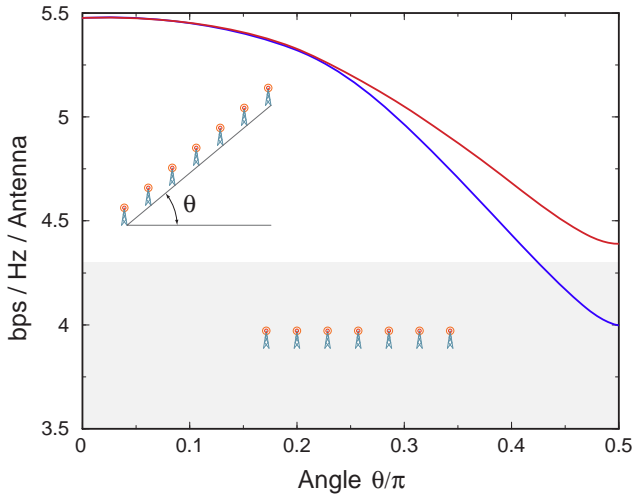}\vspace{0.1in}
\caption{Maximum information transfer rate per antenna (bits per
second per Hertz) between two 1-dimensional 50 antenna arrays with one
placed deep inside a half-infinite diffusive medium (with antenna
spacing $.5 \lambda$) and the other above it (with spacing $0.6
\lambda$) at an angle $\theta$ to the surface.  Lower curve
corresponds to the maximum transfer rate for transmission signal
covariance $Q_{ij}$ equal to $\delta_{ij}$.  Upper curve is the
information capacity, optimizing over $Q$ in Eq.~3.  The signal to
noise ratio is again $100$ and scalar waves are treated.  Because the
scattered signal approaches the antennas from only one side (the
diffusive medium), the correlations in the perpendicular direction
cannot become destructive.  As a result, the correlations can only be
minimized when all the antennas are positioned at the same height from
the surface.  Thus, the maximum $5.478$ bps/Hz/antenna occurs at
$\theta=0$, when the array is parallel to the surface dividing the
diffusive region from free space. Note that optimization over $Q$
becomes more important when $\theta$ increases and there is higher
antenna redundancy.}
\label{fig3}
\end{figure}

The simplest case in which the geometry of the environment is taken
into account is a half-infinite diffusive 3-dimensional region without
any scattering in the other half space (Fig. \ref{fig3}). This is a
first approximation to modeling electromagnetic propagation in a city
of high-rise buildings.  When either of the arrays is close to the
surface of the diffusive region, the correlations between antennas
become qualitatively different.  Specifically, when one antenna array
is outside while the other is deep inside the scattering medium,
$w(\boldhat{k}) = (\boldhat{k} \cdot \boldhat{z}) \Theta(\boldhat{k}
\cdot \boldhat{z} )$ for the outside antenna array (for either scalar
or vector waves) where $\boldhat{z}$ is the normal to the interface
and $\Theta$ is the Heaviside step function.  Surprisingly, the
maximum capacity is achieved by pointing the transmitter directly into
the diffusive medium rather than at the receiver!

We close by pointing out several simple extensions which make these methods
applicable to a wide variety of realistic situations.  First, more
realistic modeling of the scattering medium could take into account
scatterers that are not small compared to the wavelength. A concrete
example is the scattering off walls inside a building where the
scattering is highly anisotropic {\em (9)}. This case can
be analyzed within our framework; however, the correlations no longer
have the simple factorized form Eq. \ref{RST}.  

Second, more realistic antennas can be treated by simply inserting
appropriate functions $\chi_i (\boldhat{k},\boldhat{\epsilon})$ in Eq.
\ref{chichi}.  Even for nontrivial antennas, these functions can be
either determined numerically or measured experimentally. In this way,
both practical antenna designs and the effects of antenna interactions
can be included.

Finally, we can consider the case where the transmitter also has
knowledge of $G$.  In this case, the capacity will be greater. Here,
the averaging over $G$ occurs after optimizing over $Q$, a
process known as ``water-filling'' {\em (2)}.  This
case can also be handled within our framework.



\begin{references}
\bibitem{} Formally, we pose this information theory problem
  as follows.  The received signal vector is $y = G x + \eta$ where
  $x$ is the transmitted vector and $\eta$ is the noise vector.  Thus,
  the channel is defined by $p(\{G, y\}|x) = p(G) p(y| \{G, x\})$
  where $ p(G) \propto \exp( - \case{n_T}{2 S} {\rm{Tr}} [G T^{-1}
  G^\dagger R^{-1}]) $ gives $G$ the proper statistics as defined by
  Eq.  \ref{RST}, and $ p(y | \{ G, x\} ) \propto \exp( - \case{1}{2N}
  |y - Gx|^2) $ describes independent noise of average power $N$ in
  each component of $\eta$.  The capacity as shown in Eq.
  \ref{matrixshannon} is then the mutual information $I(\{G, y \}; x)$
  maximized over all possible $p(x)$ transmitted.
\end{references}
\end{document}